\documentclass[%
preprint,
nofootinbib,
amsmath,
amssymb,
aps,
prd,
]{revtex4-2}

\usepackage{graphicx}
\usepackage{dcolumn}
\usepackage{bm}
\usepackage{float} 
\usepackage{comment} 
\usepackage{layouts} 
\usepackage{nicefrac} 
\usepackage{empheq} 
\usepackage{hyperref}
\usepackage[mathlines]{lineno}

\usepackage{xspace}
\usepackage{xcolor}

\newcommand{\F}{\textit{Fermi}\xspace}
\newcommand{\di}{\mathrm{d}}

\begin{document}


\title{Constraints on the antistar fraction in the Solar system neighborhood from the 10-years \F Large Area Telescope gamma-ray source catalog}

\author{Simon Dupourqué}
\author{Luigi Tibaldo}
\author{Peter von Ballmoos}
\affiliation{%
Institut de Recherche en Astrophysique et Planétologie (IRAP)\\
Université de Toulouse, CNRS, UPS, CNES\\
31400 Toulouse, France
}%

\date{\today}

\begin{abstract}
\end{abstract}

\maketitle


\section{\label{sec:intro}Introduction}
We generally take it for granted that equal amounts of matter and antimatter were produced in the Big Bang, yet the observable Universe seems to contain only negligible quantities of antimatter. Baryonic antimatter in our Solar and Galactic neighborhood can be constrained by the observation of high-energy gamma rays \cite{Steigman1976}: when coming into contact with normal matter, it would produce annihilation radiation featuring a characteristic spectrum peaking around half the mass of the neutral pion at $\sim$70~MeV, and with a cutoff around the mass of the proton at 938~MeV \cite{Backentoss1984}. The non detection of this annihilation feature in gamma rays has virtually excluded the existence of substantial amounts of baryonic antimatter in the solar system, the solar neighborhood, the Milky Way, and up to the scale of galaxy clusters \cite{Steigman1976,Steigman2008,vonBallmoos2014}. When combined with observations of the largely isotropic Cosmic Microwave Background, the lack of an ``MeV-bump'' in gamma-rays has led to the presently accepted paradigm in which a matter-antimatter symmetric Universe can be ruled out \cite{Cohen1998}. 

Presently, baryon asymmetry is regarded as one of the deepest enigmas of nature. While emerging in the macroscopic Universe, its origin has been sought mainly in the microscopic world of particle physics. The discoveries that weak interactions violate parity invariance (P violation \cite{Wu1957}) and charge-parity symmetry (CP violation \cite{Christenson1964}) were the first experimental clues leading to {\it baryogenesis} scenarios for explaining the excess of matter over antimatter.
In baryogenesis scenarios, the reheating that follows the inflationary epoch produces an initially symmetric universe (equal abundances of matter and antimatter), and then departure from the CP invariant state out of thermal equilibrium and the dynamical production of a net baryon number result in the observed baryon asymmetry \cite{Sakharov1967}.

The hitherto observed symmetry violations are, however, far too minute to explain the observed baryon asymmetry quantitatively. Despite continuing efforts, baryon-number violating processes have not yet been observed, and CP violation in the quark sector seems many orders of magnitude below what the observed baryon asymmetry would require.  
Recent results from the T2K experiment indicate that CP symmetry might be violated in the lepton sector \cite{T2K2020}, pointing towards a process called {\it leptogenesis} for generating the matter–antimatter asymmetry. A primordial imbalance of the number of leptons over antileptons would be later converted to a baryon asymmetry.

While some form of baryogenesis or leptogenesis can still be considered the prevailing explanation for the observed baryon asymmetry, this is nowhere near being on a firm footing. Consequently, alternative scenarios for solving the problem are appearing - or make surface again. Amongst a long list of competing theories, let us point out only two of the more recent ones: the Dirac-Milne Universe of \citeauthor{Levy2012} \cite{Levy2012}, under scrutiny via the experimental study of the gravitational behavior of antimatter \cite{Perez2017}, and the CPT-symmetric Universe of \citeauthor{Turok2018} \cite{Turok2018}, which could explain recent tantalizing observations by the ANITA experiment  \cite{Anchordoqui2018}.

The standard paradigm that our local Universe is completely matter dominated has recently been challenged by the tentative detection of a few anti-helium nuclei by the Alpha Magnetic Spectrometer experiment (AMS-02) on the International Space Station  \cite{AMS-STing}. AMS-02 measures roughly one anti-helium in a hundred million helium. Amongst the eight anti-helium events reported, six are compatible with being anti-helium-3 and two with anti-helium-4. Several authors, e.g., \citeauthor{Salati1999} \cite{Salati1999}, had pointed out that ``the detection of a single anti-helium [\ldots] would be a smoking gun [\ldots] for the existence of antistars and of antigalaxies''.

Nevertheless, alternative explanations for the AMS-02 events have been explored by \citeauthor{poulin2019} \cite{poulin2019}. They concluded that neither spallation from primary cosmic-ray protons and helium nuclei onto the interstellar medium (ISM), nor the annihilation of hypothetical dark-matter particles seems to be able to explain the observed flux of anti-helium. They give substance to the hypothesis that the only way to account for the observation of anti-helium, if it is confirmed, is indeed the existence of nearby anticlouds or antistars, with the most likely explanation given by antistars in the solar neighborhood. While the latest efforts rule out even more convincingly the spallation hypothesis \cite{shukla2020}, the tuning of dark-matter theories to produce larger quantities of anti-nuclei is still an open avenue \cite{heeck2020,cholis2020}. For a recent review on the subject see also \citeauthor{antinreview2020} \cite{antinreview2020}.

As discussed since \citet{Steigman1976} and recently remarked by \citeauthor{poulin2019} \cite{poulin2019}, gamma-ray observations can be used to constrain the abundance of nearby antistars. Therefore, in this paper we use the recently published 4\textsuperscript{th} catalog of high-energy gamma-ray sources detected with the \F~Large Area Telescope (LAT) data release 2 (4FGL-DR2) \cite{4FGL,4FGL-DR2} to derive constraints on the existence of antistars in the solar neighborhood. The paper is organized as follows: in Section~\ref{sec:4FGL} we select antistar candidates in 4FGL-DR2 and compute the sensitivity of the LAT to an antistar signal; in Section~\ref{sec:fraction} we use the 4FGL-DR2 candidates and the sensitivity we determined to constrain the antistar fraction using various methods and assumptions; finally Section~\ref{sec:conclusions} presents a summary of our work and some discussions on its implications and future perspectives.

\section{\label{sec:4FGL}Constraining antistars with 4FGL-DR2}

\subsection{\label{sec:candidates}Antistar candidates in 4FGL-DR2}

4FGL-DR2 \cite{4FGL,4FGL-DR2} is based on 10 years of observations with the LAT in the energy range from 50 MeV to 1 TeV. It contains 5787 gamma-ray sources with their spectral parameters, spectral energy distributions, light curves, and multiwavelength associations.
Source detection in 4FGL-DR2 is based on the likelihood ratio test. More specifically it is based on the Test Statistic (TS) defined as  
\begin{equation}
    \text{TS} = 2 \log \frac{\mathcal{L}}{\mathcal{L}_0}
\end{equation}
where $\mathcal{L}$ is the likelihood of the model including the candidate gamma-ray source and $\mathcal{L}_0$ is the likelihood of the background model not including the source. The main backgrounds for source detection in the LAT band are interstellar gamma-ray emission produced by interactions of cosmic rays with interstellar matter and fields, and the isotropic background that is a mix of extragalactic diffuse emission and a residual contamination from CR interactions in the LAT misclassified as gamma rays. 

We select antistar candidates in 4FGL-DR2\footnote{We used the initial release of the catalog (file \texttt{gll\_psc\_v23.fit}), but we checked that all results are unchanged for the latest version available at the moment of writing which includes more optical classifications (file \texttt{gll\_psc\_v26.fit}).} based on the following criteria:
\begin{itemize}
    \item extended sources are excluded since the angular size of a star is several orders of magnitude smaller than the LAT resolution at low energy, thus antistars are expected to be point-like sources;
    \item sources associated with objects known from other wavelengths that belong to established gamma-ray source classes (e.g., pulsars, active galactic nuclei) are excluded;
    \item sources with total TS summed for energy bands above 1 GeV larger than 9 (that is, emission detected at $> 3\sigma$ above 1 GeV) are excluded since the emission spectrum from proton-antiproton annihilation is null above 938 MeV (mass of the proton); the high-energy cutoff makes it possible to differentiate the matter-antimatter annihilation signal from the well-known pion-bump signal produced by interactions of cosmic rays with an approximate power-law spectrum onto the ISM and seen in the Galactic interstellar emission and a few supernova remnants \cite{Ackermann2013,Jogler2016}; to our knowledge this is the first time that spectral criteria are used to select candidate antistars in gamma-ray catalogs;
    \item sources flagged in the catalog as potential spurious detections related to uncertainties in the background models or nearby bright sources (flags 1 to 6) are excluded. 
\end{itemize}

\begin{figure*}
\includegraphics{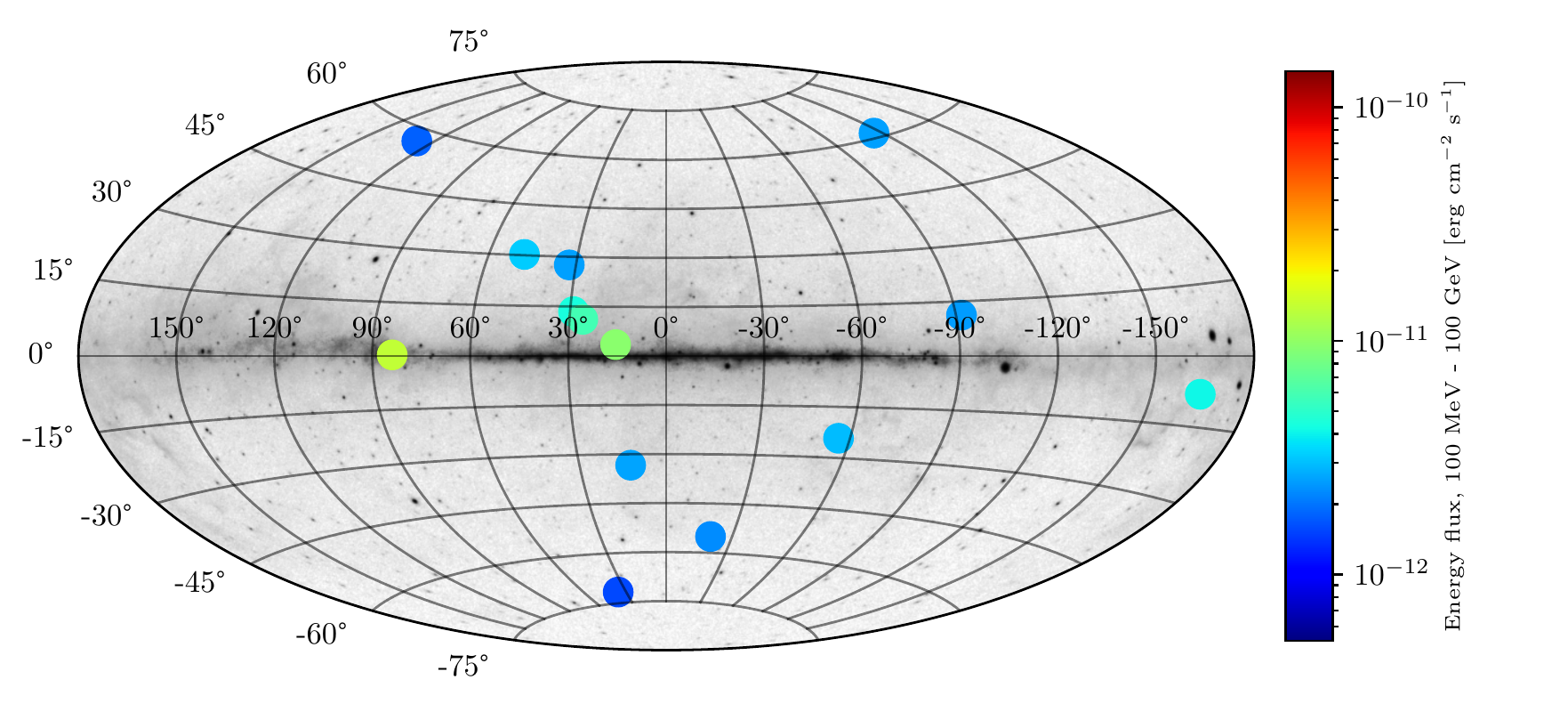}
\caption{\label{fig:sources} Positions and energy flux in the 100 MeV - 100 GeV range of antistar candidates selected in 4FGL-DR2. Galactic coordinates. The background image shows the Fermi 5-year all-sky photon counts above 1 GeV (Image credit: NASA/DOE/Fermi LAT Collaboration) 
}
\end{figure*}

This results in 14 antistar candidates listed in Table \ref{tab:sources_specs}. Figure \ref{fig:sources} shows their positions in the sky and fluxes. They do not follow a particular pattern on the sky, and they are all faint and close to the LAT detectability threshold. Therefore, their spectra\footnote{Spectra are available on the 4FGL-DR2 webpage at \url{https://fermi.gsfc.nasa.gov/ssc/data/access/lat/10yr_catalog/}.} are characterized by sizable uncertainties. The nature of these sources cannot be firmly established at present. Besides the tentative antistar interpretation, they may be sources belonging to a known gamma-ray source class, such as pulsars or active galactic nuclei, that could be identified by searching for periodicity in gamma-ray \cite[e.g.,][]{Abdo2009} and radio data \cite[e.g.,][]{Keith2011}, or for spectral signatures in optical and infrared observations \cite[e.g.,][]{Pe_a_Herazo2020}, respectively. Furthermore, they may also correspond to imperfections of the background interstellar emission model, e.g, owing to limitations of ISM tracers, for which improvements can be achieved thanks to multiwavelength data (for details on the latter aspect see, e.g., \cite{4FGL}). Identifying the sources as antistars seems more challenging, and may be attempted, for instance, using X-ray polarimetry \cite{dolgov2014pol}. Proving or disproving the antistar interpretation therefore requires significant multiwavelength work which is beyond the scope of this paper. In the following we will use the candidate list to set upper limits on the antistar abundance in the region around the Sun. 

\begin{table*}
\caption{\label{tab:sources_specs}
Antistar candidates in 4FGL-DR2 and their properties: Galactic longitude $l$, Galactic latitude $b$, energy flux $J$ and photon flux $\Phi$ in broad energy ranges given by the catalog, and $TS$ summed for the energy bands $> 1$~GeV.}
\begin{ruledtabular}
\begin{tabular}{lccccc}
 Name              & $l$ & $b$ & $J$ (0.1 - 100 GeV) & $\Phi$ (1 - 100 GeV) & TS (1 - 100 GeV)   \\
 & degrees & degrees &   erg cm$^{-2}$ s$^{-1}$ & cm$^{-2}$ s$^{-1}$ &  \\
\hline
 4FGL J0548.6+1200 &                194.9 &                -8.1 & $(4.2\pm 0.9) \times 10^{-12}$       & $(2.0\pm0.6) \times 10^{-10}$      &      8.17001  \\
 4FGL J0948.0-3859 &                268.3 &                11.2 & $(2.5\pm 0.7) \times 10^{-12}$       & $(1.4\pm0.5) \times 10^{-10}$      &      3.17782  \\
 4FGL J1112.0+1021 &                243.8 &                61.2 & $(2.5\pm 0.5) \times 10^{-12}$       & $(6.0\pm2.4) \times 10^{-11}$      &      6.18527  \\
 4FGL J1232.1+5953 &                127.4 &                57.1 & $(1.8\pm 0.3) \times 10^{-12}$       & $(4.7\pm1.8) \times 10^{-11}$      &      3.27565  \\
 4FGL J1348.5-8700 &                303.7 &               -24.2 & $(3.0\pm 0.6) \times 10^{-12}$       & $(9.3\pm3.3) \times 10^{-11}$      &      7.04146  \\
 4FGL J1710.8+1135 &                 32.2 &                27.5 & $(2.5\pm 0.6) \times 10^{-12}$       & $(4.8\pm2.3) \times 10^{-11}$      &      0.552135 \\
 4FGL J1721.4+2529 &                 48.1 &                30.2 & $(3.3\pm 0.5) \times 10^{-12}$       & $(1.1\pm0.3) \times 10^{-10}$      &      8.78427  \\
 4FGL J1756.3+0236 &                 28.9 &                13.4 & $(4.4\pm 1.0) \times 10^{-12}$       & $(1.9\pm0.5) \times 10^{-10}$      &      5.02135  \\
 4FGL J1759.0-0107 &                 25.9 &                11.1 & $(5.9\pm 1.3) \times 10^{-12}$       & $(2.5\pm0.6) \times 10^{-10}$      &      8.62541  \\
 4FGL J1806.2-1347 &                 15.5 &                 3.5 & $(9.4\pm 2.2) \times 10^{-12}$       & $(4.7\pm1.0) \times 10^{-10}$      &      7.76874  \\
 4FGL J2029.1-3050 &                 12.3 &               -33.4 & $(2.6\pm 0.6) \times 10^{-12}$       & $(1.2\pm0.4) \times 10^{-10}$      &      7.99515  \\
 4FGL J2047.5+4356 &                 83.9 &                 0.3 & $(1.4\pm 0.4) \times 10^{-11}$       & $(4.5\pm1.6) \times 10^{-10}$      &      5.17449  \\
 4FGL J2237.6-5126 &                339.8 &               -55.0 & $(2.3\pm 0.5) \times 10^{-12}$       & $(7.0\pm2.4) \times 10^{-11}$      &      0.714205 \\
 4FGL J2330.5-2445 &                 35.8 &               -71.7 & $(1.6\pm 0.4) \times 10^{-12}$       & $(8.6\pm2.6) \times 10^{-11}$      &      8.69572  \\
\end{tabular}
\end{ruledtabular}
\end{table*}

\subsection{\label{sec:sens}Sensitivity to an antistar signal}

In order to use the candidates to constrain the population of antistars in the solar neighborhood we need to establish the sensitivity to an antistar signal in 4FGL-DR2. 4FGL-DR2 sources were selected based on the criterion that $\mathrm{TS} \ge 25$ over the entire energy band from 50 MeV to 1 TeV (detection significance $\ge 4.1 \sigma$). Thus, to determine whether an antistar would appear in the catalog, we need to determine the minimum flux that it must have so that its $\mathrm{TS} = 25$.

To do so, we follow the method proposed in Appendix A of the first \F-LAT source catalog (1FGL) \cite{1FGL}. This consists in calculating semi-analytically TS for a pointlike source based on the Instrument Response Functions (IRFs) of the LAT, the spectrum of the source  $S(E)$ and the background model $B(E)$. In order to account for photons with different reconstruction qualities, the analysis in 4FGL-DR2 is separated for 15 components, detailed in Table 2 of the 4FGL article \cite{4FGL}. The analysis used to build 4FGL includes in addition weights $w(E)$ to take into account the systematic uncertainties of the background model. By extending the semi-analytical TS formula from 1FGL to account for the multiple components and weights we obtain:
\begin{equation}
\text{TS} = \sum_{\text{component}} \int^{E_{\text{max}}}_{E_{\text{min}}}E\; \di \log E\;  \left\{ 2 \, Exp(E) \, A(E) \, B(E) \, w(E) \right\}.  \label{eq:TS}
\end{equation}
where $A(E)$ is defined as 
\begin{empheq}[left = \empheqlbrace]{align*}
                &A(E) = \int^{\theta_{\text{max}}}_{0}2\pi\sin \theta \di\theta\left\{\left( 1 + g(\theta,E)\right) \log\left( 1 + g(\theta,E)\right) -  g(\theta,E)\right\}\\
               &g(\theta,E) = \frac{S(E) \, PSF(\theta,E)}{B(E)}.
\end{empheq}

Let us define the different terms involved in the equation \ref{eq:TS}: 
\begin{itemize}
\item $PSF(\theta,E)$ [sr$^{-1}$] represents the Point Spread Function (PSF) of the LAT, as a function of $\theta$, the angular distance to the source position, and $E$, the energy of the photon;
\item $Exp(E)$ [cm$^2$ s] represents the exposure, i.e. the product of the effective area and the observation livetime;
\item $B(E)$ [MeV$^{-1}$ cm$^{-2}$ s$^{-1}$ sr$^{-1}$] represents the interstellar and isotropic background model.
\item $S(E)$ [MeV$^{-1}$ cm$^{-2}$ s$^{-1}$] represents the spectrum of the source;
\item $w(E)$ represents the weights introduced into the analysis of 4FGL.
\end{itemize}

The exposure, PSF, and background intensity in formula \ref{eq:TS} vary as function of position in the sky. We calculated the first two for the list of Good Time Intervals and set of IRFs used in 4FGL-DR2 using the \texttt{fermitools} version~1.2.23. We also employ the background models from 4FGL-DR2 (see \citeauthor{gll_iem} \cite{gll_iem} for more details on the methodology to construct the model). All maps are calculated in Galactic coordinates with a resolution of $(0.125^\circ)^2$, corresponding to the resolution of the background model, and in Hammer-Aitoff projection in order to minimize distortions at high latitudes. 
The source spectrum $S(E)$ is assumed to be the $p-\overline{p}$ annihilation spectrum from \citeauthor{Backentoss1984} \cite{Backentoss1984}.
In order to approximately account for the source confusion limit, the solid angle integral is computed up to the mean angular distance between sources in the catalog in 4FGL-DR2 $\theta_\text{max} = 1.5062 \deg$ \cite{1FGL}. The weights are calculated according to Appendix~B in the 4FGL paper \cite{4FGL}, which requires to calculate the number of background events within the PSF of the LAT for each energy band. We use model-based weights derived from the 4FGL-DR2 background model. The number of counts $N_{k}$ from Equation~B.4 of 4FGL thus becomes:
\begin{equation}
N_{k}(E) = \int_{E}^{2E} \di E' Exp(E')B(E') \int_0^{\theta_\text{max}} 2\pi \sin \theta \di \theta \frac{PSF(\theta,E')}{PSF(0,E')}.
\end{equation}

\begin{figure*}
\includegraphics[width=\textwidth]{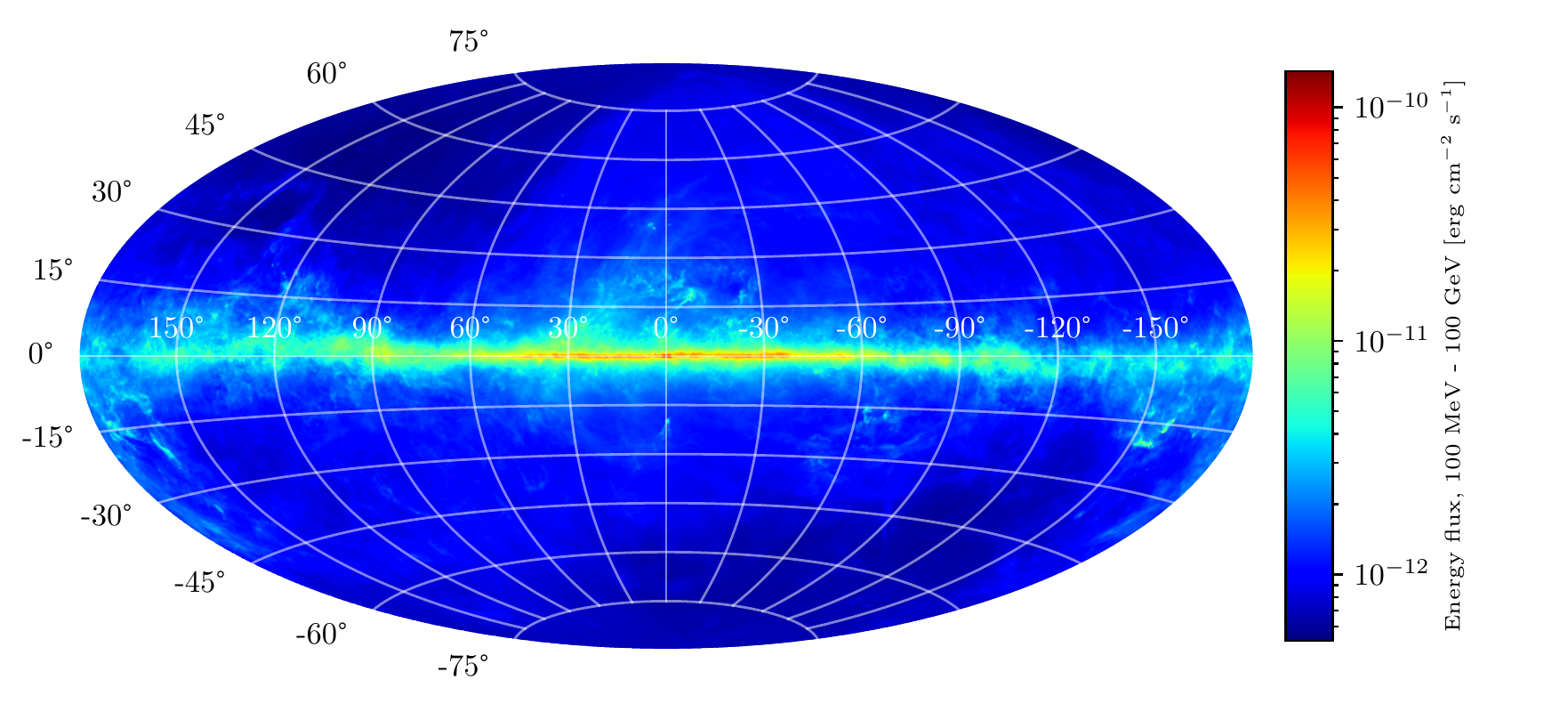}
\caption{\label{fig:sensitivity} Minimum energy flux in the 100 MeV - 100 GeV energy range for a pointlike source with matter-antimatter annihilation spectrum to be detectable in the 4FGL-DR2 catalog. Galactic coordinates.}
\end{figure*}

The LAT sensitivity to an antistar signal can therefore be expressed in the form of a sky map, where each pixel represents the flux necessary to obtain TS = 25 for a pointlike source with a matter-antimatter annihilation spectrum at this position. The resulting sky map is shown in Figure~\ref{fig:sensitivity} and  also available in machine-readable format at the CDS\footnote{Available at CDS via anonymous ftp to cdsarc.u-strasbg.fr (130.79.128.5) or via \url{http://cdsarc.u-strasbg.fr/viz-bin/qcat?J/other/PhRvD}}. It is given in units of energy flux integrated in the energy range from 100~MeV to 100~GeV to be readily comparable to 4FGL-DR2.
Since the main background is given by Galactic interstellar emission, as expected antistars would be more easily observed outside the Galactic plane, which tends to be the case for our candidates.

Our estimate of the sensitivity is not fully consistent with the analysis used to build the 4FGL catalog because the $p-\overline{p}$ spectrum is not among the spectral forms considered for source detection. We calculated the sensitivity for a pointlike source with a power-law spectrum of spectral index 2.7, which is used for the detection of soft sources in 4FGL \citep[][Table 3]{4FGL}. This does not entirely match the case of interest either, i.e., a source with $p-\overline{p}$ annihilation spectrum analysed by assuming a power-law spectrum. However, we can use the result to gauge the impact on our limits on antistars. The sensitivity for a power-law source of spectral index 2.7 is always better than for the $p-\overline{p}$ annihilation spectrum, with a median ratio over the sky for the minimum detectable energy flux in the 100 MeV - 100 GeV energy range of 0.67. For the rest of the paper we will use the more conservative estimate of the sensitivity based on the $p-\overline{p}$ annihilation spectrum. Using the sensitivity for a soft power-law source would make all our limits stronger.

\section{\label{sec:fraction}The fraction of antistars in the solar system neighborhood}

\subsection{Gamma-ray flux of an antistar}
The limits on the antistar population in the solar neighborhood are established based on the hypothesis that antistars in the Galaxy would accrete matter from the ISM with subsequent $p-\overline{p}$ annihilation at their surface \cite{Steigman1976}.

Following the steps of \citeauthor{Steigman1976} \cite{Steigman1976}, we compute the total luminosity of an antistar for Bondi-Hoyle-Littleton accretion \cite{Edgar2004} and using the gamma-ray yield per $p-\overline{p}$ annihilation from \citeauthor{Backentoss1984} \cite{Backentoss1984}. Taking into account explicitly the speed of sound $c$ and the density of matter $\rho$ in the ISM, this yields
\begin{equation}
    L_\gamma = 8.45\times 10^{35} \left( \frac{\rho}{\text{ m$_p$ cm}^{-3}}\right) \left( \frac{M}{M_\odot} \right)^2 \left( \frac{\sqrt{v^2 + c^2}}{10 \text{ km\:s}^{-1}} \right)^{-3} [\text{ph s}^{-1}].
    \label{eq:lgamma}
\end{equation}
The remaining parameters are  the antistar mass $M$ and its velocity $v$ with respect to the ISM. Assuming isotropic gamma-ray emission and that there is no significant absorption during the propagation, the total source flux at a distance $d$ is $\Phi = L_\gamma/4\pi d^2$. 

Owing to the unavailability of measurements of the annihilation cross sections for reactions of antinuclei other than $p-\overline{p}$ and the lack of robust prescriptions on the elemental and isotopic composition of antistars, all along this study we neglect the effect of species heavier than $p$ both in antistars and in the ISM. Taking those into account would make all the upper limits derived in the following sections stronger.

Beside antistar properties, the calculation of the gamma-ray fluxes requires some knowledge about the ISM.
\begin{itemize}
\item Throughout this work we fix $c = 1$ km s$^{-1}$, i.e., the isothermal sound speed of the dominant cool atomic phase in the ISM at a temperature of 100 K \cite{draine2011}. Variations of $c$ of a factor of a few that are known to occur in the ISM are not expected to change substantially our conclusions for antistars with velocities ranging from tens to hundreds of km s$^{-1}$ which will be mainly discussed below. 
\item In Sections~\ref{subsec:starlike} and~\ref{subsec:primordial} the density of interstellar hydrogen at the antistar positions is calculated based on the model by \citeauthor{Shibata2011} \cite{Shibata2011}.
\item In Sections~\ref{subsec:starlike} and~\ref{subsec:primordial} the velocity of the antistars is converted into velocity with respect to the ISM under the hypothesis of purely circular motion of the ISM around the Galactic center, described by the universal rotation curve of \citeauthor{Persic1996} \cite{Persic1996} with the parameters for the Milky Way inferred from recent parallax distance measurements of high-mass star-forming regions \cite{reid2019}. 
\end{itemize}

\subsection{Parametric derivation of the antistar fraction}
In this section we establish limits on the antistar fraction based on the method proposed by \citeauthor{Steigman1976} \cite{Steigman1976} and largely employed in the earlier literature on the subject. The method consists in assuming that the brightest antistar candidate is the nearest antistar. One can thus determine its distance based on its photon flux $\Phi_{\max}$ for any mass, velocity, and ISM density values. The sphere with radius equal to such distance is assumed to contain at most one antistar, and the fraction of antistars to normal stars is given by $f_{\bar{\ast}} = (n_\ast V )^{-1}$ where $n_\ast$ is the local star density (for which we assume the value of $0.15 \text{ pc}^{-3}$ from \citeauthor{Latyshev1978} \cite{Latyshev1978}), and $V$ is the volume of the sphere.

In parametric form, the antistar fraction upper limit is given by 
\begin{equation}
    f_{\bar{\ast}} \leq 2.68 \times 10^3 \left(\frac{ \Phi_{\max}}{\text{ cm}^{-2}\text{ s}^{-1}}\right)^{3/2}\left( \frac{\rho}{\text{ m$_p$ cm}^{-3}}\right)^{-3/2}\left(\frac{M}{M_\odot}\right)^{-3}\left(\frac{\sqrt{v^2 + c^2}}{10\text{ km}\text{ s}^{-1}} \right)^{9/2}
    \label{eq:steig1}
\end{equation}

\begin{figure*}
\includegraphics{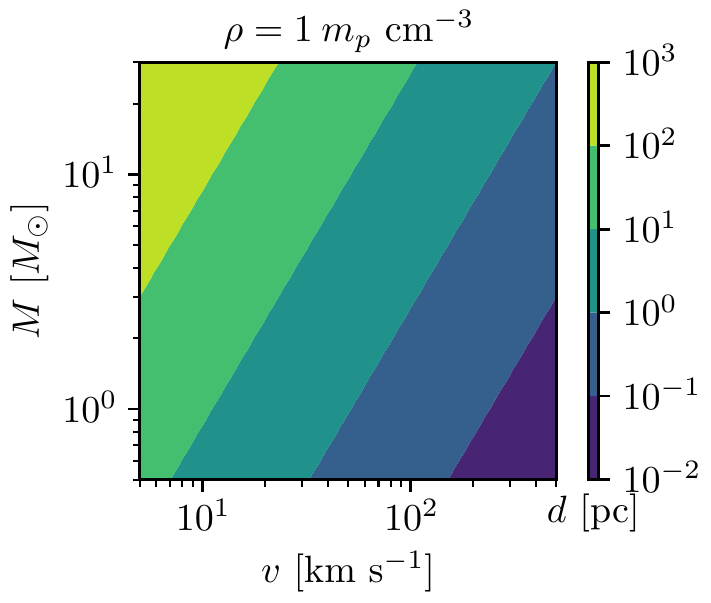}
\includegraphics{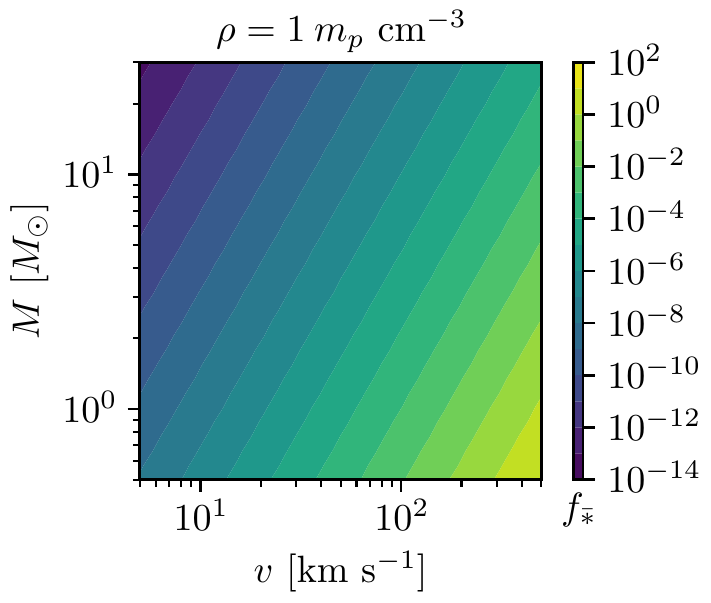}
\caption{\label{fig:dmin}
Left: distance $d$ of the closest antistar candidate in the 4FGL-DR2 catalog based on the luminosity relation in Equation~\ref{eq:lgamma}. Right: corresponding upper limit on the antistar fraction $f_{\bar{\ast}}$ from Equation~\ref{eq:steig1}. In both panels the quantities are shown as a function of velocity $v$ w.r.t. the ISM and antistar mass $M$ for an ISM density $\rho = 1$~$m_p$~cm$^{-3}$.
}
\end{figure*}

We use the energy flux in the 100~MeV-100~GeV energy range from 4FGL-DR2 to obtain the total photon flux for the $p-\overline{p}$ annihilation spectrum \cite{Backentoss1984}, and thus obtain the minimal distances and upper limits on the antistar fraction shown in in Figure~\ref{fig:dmin} as a function of antistar mass and velocity for an ISM density of $\rho = 1$~m$_p$~cm$^{-3}$.

The distance to the closest antistar and corresponding antistar fraction varies very much based on the assumed parameter values.
For example, taking $M = 1\;M_\odot$, $v = 10$~km~s$^{-1}$, and $\rho = 1$~m$_p$~cm$^{-3}$ the closest antistar would be at 10~pc, which would yield an upper limit on the fraction $f_{\bar{\ast}} \leq 10^{-8}$. For comparison, the upper limit provided by \citeauthor{Steigman1976} was $f_{\bar{\ast}} \leq 10^{-4}$ based on SAS-2 data \cite{Steigman1976}. 

In 2014, \citeauthor{vonBallmoos2014} inferred an upper limit $f_{\bar{\ast}} < 4 \times 10^{-5}$ using unassociated sources from the LAT 2-year Source Catalog 2FGL \cite{vonBallmoos2014}. Our upper limit on the antistar fraction is stronger because antistar candidates are selected according to more restrictive criteria, notably the lack of significant emission above 1 GeV, drastically reducing their number. Moreover, the accumulation of additional data by the LAT also makes it possible to observe sources whose photon flux is 10 times lower than those selected by \citeauthor{vonBallmoos2014} \cite{vonBallmoos2014}: their distance would be larger, thus lowering the upper limit on the fraction.

This method for estimating $f_{\bar{\ast}}$ has several limitations: it relies on arbitrary choices for the parameters and the obtained limits do not have a well-defined statistical meaning. In addition, Equation~\ref{eq:steig1} takes into account the flux of one source only, neglecting the rest of the exploitable information. 

\subsection{Monte Carlo derivation of the antistar fraction}

\subsubsection{\label{sec:MCmethod}General method description}

To overcome the limitations of the previous procedure, we propose a novel Monte~Carlo method. The method relies on a well-defined hypothesis on the antistar population with only one free parameter (the antistar fraction $f_{\bar{\ast}}$ or the antistar density $n_{\bar{\ast}}$). Based on this hypothesis we build an estimator $\hat{N}_{\bar{\ast}}$ for the number of antistars  that should be detected for a given value of the free parameter. For each parameter value, we generate 1000 synthetic antistar populations according to the hypothesis and calculate the associated gamma-ray fluxes. The fluxes are then compared to the sensitivity map (Figure~\ref{fig:sensitivity}) to check whether the synthetic sources would be detected or not, and determine the number of expected detections.

We note that this method does not provide accurate results w.r.t. to effects relevant to individual sources (e.g., presence of a nearby source, small-scale fluctuation of the ISM density). However, owing to the large number of populations generated, the procedure should provide a reliable estimation of the average number of expected detections. 

We determine the value of the free parameters that yields $\hat{N}_{\bar{\ast}} \leq 14$ for 95\% of the synthetic populations, and $\hat{N}_{\bar{\ast}} > 14$ for 5\% of the populations, where 14 is the number of antistar candidates found in 4FGL-DR2 (Section~\ref{sec:candidates}). This provides a 95\% confidence level upper limit on the parameter value. The value is determined via the probabilistic bisection algorithm \cite{PBA}, which is an adaptation of the classical bisection algorithm for stochastic root finding.
We use the implementation\footnote{Publicly available at \url{https://github.com/choderalab/thresholds}.} of this algorithm by \citeauthor{Fass2018} \cite{Fass2018} with the maximum number of iterations on the parameter value set to 1000.

\subsubsection{\label{subsec:starlike}Antistar fraction for a young disk population}
The first hypothesis that we consider is that antistars have the same properties as normal stars, dominated by the young stellar populations in the Galactic disk. Although difficult to justify physically\footnote{See, e.g., the discussion in \citeauthor{poulin2019} \cite{poulin2019} on the challenges to the hypothesis that antistars are actively forming in the Milky Way at the current epoch, which would require the survival of anticlouds from the early Universe within the Galactic ISM.}, this hypothesis makes it possible to compare our results with previous works that employ star-like parameters for antistars. In order to generate synthetic star population we use the code \texttt{Galaxia} \cite{Galaxia}, which implements the state-of-the-art Besançon model \cite{besanconRef}. The free parameter here is the fraction $f_{\bar{\ast}}$, and the generation of a Monte Carlo population at a given $f_{\bar{\ast}}$ is done by randomly selecting stars from a \texttt{Galaxia} population with probability $f_{\bar{\ast}}$. Populations are generated for a maximum distance of 11 kpc, which corresponds to the maximal detection distance by the LAT for star properties according to the considered model.

Using this method, the local fraction of antistars is estimated at $f_{\bar{\ast}} < 2.5 \times 10^{-6}$ at 95\% confidence level. This result is 20 times more constraining than the limit reported based on 2FGL \cite{vonBallmoos2014}, and no longer relies on arbitrary choices for the antistar properties.

\begin{figure*}
\includegraphics[width=\textwidth]{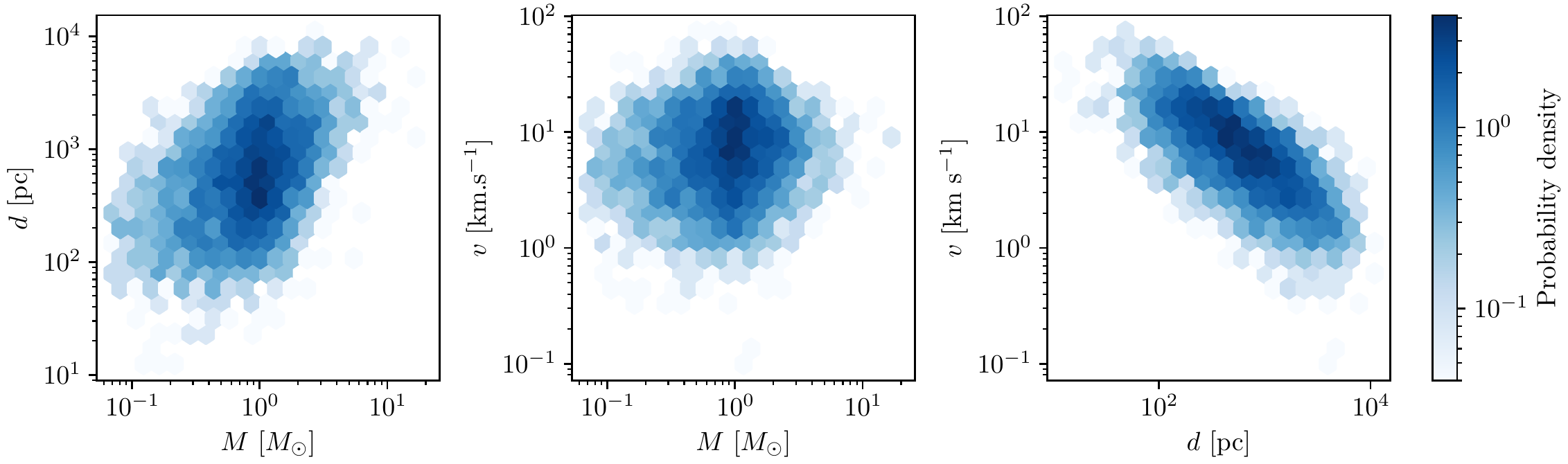}
\caption{\label{fig:starplan} Distribution of detected-antistar properties obtained from a \texttt{Galaxia} synthetic population, by comparing the flux from Equation \ref{eq:lgamma} with the LAT sensitivity (see Figure \ref{fig:sensitivity}).
}
\end{figure*}

For starlike properties the antistars more likely detected by the LAT would have masses $\sim$1~$M_\odot$, velocities w.r.t. the ISM of $\sim$10~km~s$^{-1}$, and distances of $\sim$500~pc. Figure~\ref{fig:starplan} shows the projections of the detection probability density as a function of mass, distance, and velocity.
Stars with masses $<1\: M_\odot$, although more abundant, are not efficient enough in terms of accretion to be predominant in the detected sources. Velocity is a key component, it is mainly the speed of an antistar that will determine the maximum distance at which it can be observed. Stars of high mass and especially of low velocity are the most distant objects observable by the LAT, up to 10 kpc. 

\subsubsection{\label{subsec:primordial}Antistar fraction for a primordial halo population}

A more physically-motivated scenario discussed in the literature is that antistars may be primordial objects produced in the early Universe, e.g., in the Affleck-Dine scenario for baryogenesis \cite{dolgov2014,Blinnikov2015}. Under this hypothesis, antistars would now be present in galactic halos as a subclass of primordial baryon-dense objects (BDOs). The contribution of BDOs to halo masses was constrained so far through microlensing observations, most recently in the Magellanic Clouds by MACHO \cite{MACHO2000}, EROS \cite{EROS1997}, and EROS-2 \cite{EROS22007,OGLE2013}.

We test this scenario against the 4FGL-DR2 antistar candidates by using our Monte~Carlo method. \citeauthor{Blinnikov2015} \cite{Blinnikov2015} provide a typical velocity for primordial antistars of 500 km s$^{-1}$. Since this velocity is close to the Galactic escape velocity we expect the gravitational potential of the Milky Way to have little impact on the spatial and velocity distribution of these objects. Therefore we generate mock antistar populations with uniform spatial distribution and a velocity of 500 km s$^{-1}$ with isotropic distribution. As we lack clear model prescriptions for the mass distribution, we repeat the procedure several times for fixed mass values in the range from $0.3\; M_\odot$ to $10\; M_\odot$. The mass range is chosen to compare with earlier results from microlensing. The lower bound is driven by computational efficiency owing to the fact that a huge number of antistars is needed to reach 14 LAT detection for such low masses. As we will see in this low mass range the gamma-ray constraints are anyway weaker than other existing upper limits from microlensing. The upper bound reflects the model prediction that antistars heavier than a few solar masses are less likely to be found \cite{dolgov2014}.

The populations are generated in a sphere with a radius of 70 pc centered at the Sun position, which corresponds to the maximum distance for a detection with the LAT for objects of 10 $M_\odot$ and speed of 500 km~s$^{-1}$. For a given mass $M$, the number density of antistars $n_{\bar{\ast}}$ is determined using the procedure detailed in Section~\ref{sec:MCmethod}. The fraction of antistars $f_{\bar{\ast}}$ can then be estimated by taking the ratio of the density $n_{\bar{\ast}}$ over the local star density $n_{\ast} = 0.15 \text{ pc}^{-3}$ \cite{Latyshev1978}. We also compute the mass fraction defined as the ratio of the antistar mass to the local dark-matter density ($\rho_{\text{DM}} = 0.0088 \; M_\odot \text{ pc} ^{-3}$ \cite{2020MNRAS}),  $f_M = M n_{\bar{\ast}}/\rho_{\text{DM}}$, in order to compare with earlier results from microlensing in the Magellanic Clouds. 

\begin{figure*}
\includegraphics[width=\textwidth]{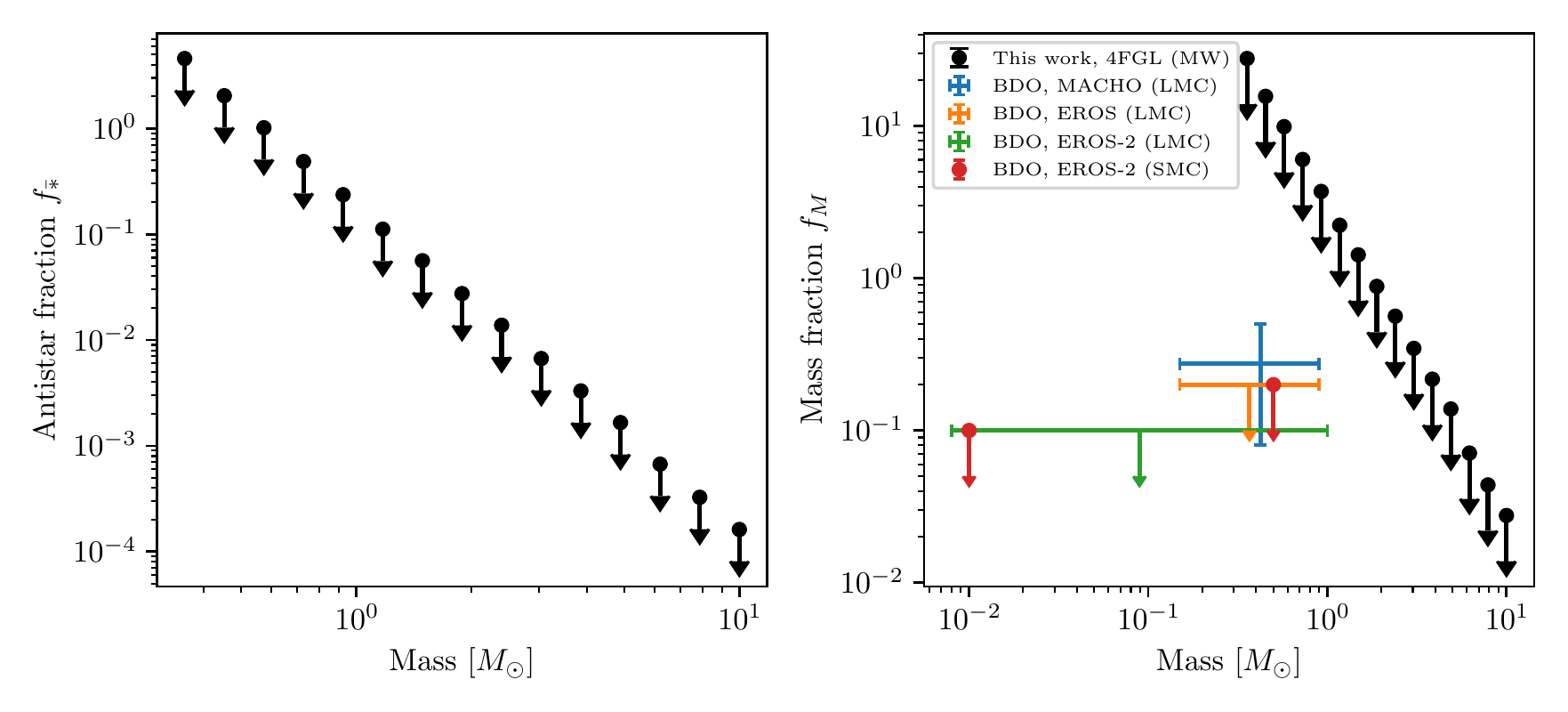}
\caption{\label{fig:macho} Antistar fraction (left) and mass fraction (right) upper limits as a function of antistar mass obtained from 4FGL-DR2 under the hypothesis that antistars are primordial objects evolving in the Milky Way (MW) halo (see text for details). In the right panel we compare gamma-ray constraints on antistars to earlier results from microlensing experiments on BDOs for the Large Magellanic Cloud (LMC) and Small Magellanic Cloud (SMC).}
\end{figure*}

The resulting antistar fraction $f_{\bar{\ast}}$ and mass fraction $f_{M}$ are shown in Figure \ref{fig:macho}. Gamma rays provide novel constraints for the mass range $\gtrsim 1 \; M_\odot$, while for smaller masses microlensing constraints from nearby galaxies remain stronger\footnote{We note that microlensing constraints apply to BDOs, while the gamma-ray constraints we derived only to antistars, which are a subclass of BDOs.}. Owing to their large speed, and therefore low accretion rates, primordial antistars can be detected by the LAT only at very limited distances from the Sun ($d<70$~pc),
and for low masses the gamma-ray upper limits are so weak that they exceed the number of observed stars and even total mass density. 

\section{\label{sec:conclusions}Summary and discussion}

We have identified in the 10-year \F~LAT gamma-ray source catalog (4FGL-DR2) 14 antistar candidates that are not associated with any objects belonging to established gamma-ray source classes and are spectrally compatible with the expected signal from baryon-antibaryon annihilation. Furthermore, we have calculated the sensitivity of 4FGL-DR2 to pointlike sources powered by matter-antimatter annihilation.

Under the hypothesis that antistars in the Milky Way produce gamma rays by accreting interstellar matter that annihilates at their surface, the above results can be used to constrain the properties of hypothetical antistar populations in the Milky Way. Following the methodology used in the earlier literature on the subject, we have derived a parametric formula which provides upper limits on the antistar fraction as a function of the closest antistar mass, velocity, and surrounding medium density (Equation~\ref{eq:steig1}). Our work provides stronger upper limits than those already available thanks to the improved sensitivity reached in 4FGL-DR2, and owing to more restrictive criteria in the candidate antistar selection taking into account spectral properties.

Furthermore, we have developed a novel Monte~Carlo method that makes it possible to derive upper limits on the antistar fraction in the Solar system neighborhood (a few tens of pc to a few kpc depending on the scenario considered). It takes into account the entire sample of candidate antistars, it is based on well-defined hypotheses on the putative antistar population rather than somewhat arbitrary parameter value choices, and it provides upper limits with a well-defined statistical meaning. For an antistar population with properties equivalent to those of regular stars, dominated by the young stellar populations in the Galactic disk, the local fraction of antistars over normal stars is constrained to be $f_{\bar{\ast}} < 2.5 \times 10^{-6}$ at 95\% confidence level. This limit is $\sim$20 times more constraining than previous results based on similar hypotheses \cite{vonBallmoos2014}. For the more physically-grounded hypothesis of a primordial population of antistars in the Galactic halo, gamma rays provide new constraints for the mass range $\gtrsim 1 \; M_\odot$: the upper limits on the local antistar fraction decrease as a function of antistar mass $M$ from  $f_{\bar{\ast}} < 0.2$ at 95\% confidence level for $M = 1 \; M_\odot$ to $f_{\bar{\ast}} < 1.6 \times 10^{-4}$ at 95\% confidence level for $M = 10 \; M_\odot$.  For smaller masses microlensing constraints in the Magellanic Clouds remain stronger.

While one single antistar in the neighborhood of the Solar system might be at the origin of the anti-helium nuclei tentatively detected with AMS-02, our results strengthen earlier conclusions that a region of size $\mathcal{O}(1\;\text{pc})$ around the solar system should be free of antistars \cite{poulin2019}. The antistar-free region can be as big as $\mathcal{O}(100\;\text{pc})$ for the hypothesis that antistars share the same properties as the normal stellar population concentrated in the Galactic disk. Therefore, a population of antistars producing collectively the anti-helium seems a more likely hypothesis. Interestingly, our local gamma-ray constraints and microlensing constraints for the Magellanic Clouds extrapolated to the Milky Way in its entirety still allow the existence of a primordial population of antistars with masses $< 10\;M_\odot$ and densities lower than $\mathcal{O}(10^{-5}\;\text{pc}^{-3})$ to $\mathcal{O}(10^{-2}\;\text{pc}^{-3})$ in the Galactic halo.

The antistars more likely to be detected by the LAT lie at distances between a few tens of pc to $\sim$1~kpc, and therefore could be the same antistars producing the anti-He nuclei tentatively detected by AMS-02. However, translating quantitatively the constraints on the antistar populations to the processes at on origin of the AMS-02 anti-helium would require making hypotheses on the mechanisms that eject and accelerate anti-nuclei from antistars. Some interesting avenues are outlined by \citeauthor{poulin2019} \cite{poulin2019}: asteroid-antistar collisions, acceleration phenomena taking place in antistar clusters, and antistar-white dwarf binary mergers. Investigating this kind of hypothesis and the uncertainties in the relevant parameters and deriving combined constraints from charged species and gamma rays is beyond the scope of our paper and left to future work, but we remark that our constraints on the antistar fraction can be used to inform the development of such upcoming modeling efforts.

At the same time the gamma-ray constraints can be improved thanks to multiwavelength work to clarify the nature of antistar candidates, as well as to more sensitive surveys in gamma rays based on the continuation of the \F mission or data from a future gamma-ray telescope optimized for the MeV to GeV energy range \cite{deangelis2017,mcenery2019}.

\begin{acknowledgments}
This work was supported by the French Space Agency CNES. It is based on observations with the Large Area Telescope (LAT) embarked on the \textit {Fermi Gamma-ray Space Telescope} \cite{FERMILAT2009} and has made use of the following publicly-available software packages: numpy \cite{numpy}, matplotlib \cite{hunter2007}, Dask (\url{https://dask.org/}), and astropy \cite{astropy2013,astropy2018}. It has also made use of the NASA's Astrophysics Data System Bibliographic Services. The authors wish to thank J. Ballet for the explanations and remarks on the evaluation of the LAT sensitivity.
\end{acknowledgments}

\bibliography{references}
\end{document}